\documentclass[aps,prb,amsmath,twocolumn]{revtex4}
\usepackage{bm}
\usepackage{graphicx}
\begin{document}
\title{Statistics of current fluctuations at non-zero frequencies}

\author{Artem V. Galaktionov$^{1,3}$, Dmitri S. Golubev$^{2,3}$, and
Andrei D. Zaikin$^{1,3}$}
\affiliation{$^{1}$Forschungszentrum Karlsruhe, Institut f\"ur Nanotechnologie,
76021, Karlsruhe, Germany\\
$^2$Institut f\"ur Theoretische Festk\"orperphysik,
Universit\"at Karlsruhe, 76128 Karlsruhe, Germany \\
$^{3}$I.E. Tamm Department of Theoretical Physics, P.N.
Lebedev Physics Institute, 119991 Moscow, Russia}

\begin{abstract}
We formulate a general approach which describes statistics of
current fluctuations in mesoscopic coherent
conductors at arbitrary frequencies and in the presence of
interactions. Applying this approach to the non-interacting case, we
analyze frequency dispersion of the third cumulant of the
current operator ${\cal S}_3$ at frequencies well below both
the inverse charge relaxation time and the inverse electron dwell
time. This dispersion turns out to be important in the frequency
range comparable to applied voltages. For comparatively transparent conductors
it may lead to the sign change
of ${\cal S}_3$.

\end{abstract}
\maketitle

Investigations of current fluctuations in mesoscopic conductors can
provide a great deal of information about properties of such systems.
During last years much attention has been devoted to shot noise
\cite{bb}. Experimental and theoretical
studies of the second moment of the current operator
describing shot noise revealed a rich variety of properties caused by
an interplay between scattering, quantum coherence and charge discreteness.
Furthermore, such studies can substantially deepen our understanding
of the role of electron-electron interactions in mesoscopic transport
because shot noise and interaction effects are known to be closely
related \cite{GZ00,Levy}.

One can also go beyond the second moment and study higher
order correlators of the current operator thereby extending the amount
of information already obtained from investigations of electron transport
and shot noise. Recently the first experimental study
of the third current cumulant in mesoscopic tunnel
junctions was reported \cite{reul}.

A theoretical framework which enables one to analyze
statistics of charge transfer in mesoscopic conductors
was developed in Ref. \onlinecite{lev1}.
This theory of full counting statistics (FCS) allows to evaluate any
cumulant of the current operator in the absence of interactions and
in the zero frequency limit. Under these conditions higher order current
cumulants were investigated by a number of authors
\cite{LLY,lev2,Nag,GG}. In order to include interactions and to
analyze frequency dispersion of current fluctuations it is
necessary to go beyond the FCS theory and to develop a more general
real time path integral technique \cite{GZ00,GGZ}.

The goal of the present paper is to address statistics of current
fluctuations at non-zero frequencies. We first present a general
and formally exact expression for the real time effective action of a
coherent conductor described by an arbitrary energy independent
scattering matrix. This expression enables one to fully describe
interaction effects in such type of conductors. We will then
demonstrate that in a non-interacting case this effective action
provides a direct generalization of the FCS
generating function \cite{lev1} to non-zero frequencies.
With the aid of our technique we
will analyze the frequency dispersion of the third cumulant of
the current operator in mesoscopic coherent conductors.

It is worthwhile to point out that the frequency
dependence of current correlators can be caused by various reasons.
One of them is the effect of an external electromagnetic environment
which is important for quantitative interpretation
of the experimentally detected behavior of higher cumulants
\cite{been}. Another source of the frequency dispersion is the
internal dynamics of a quantum scatterer. Here the important
time scales are the corresponding $RC$-time $\tau_{RC}$ and the electron
dwell time $\tau_D$ inside the conductor. The latter scale was
recently taken into consideration in the analysis
of the second \cite{GZ03} and the third
\cite{nag} current cumulants for chaotic quantum dots.

In this
paper we will address current fluctuations at frequencies not
directly related to any of such scales. We will demonstrate that apart from
the above mechanisms there exists an additional -- intrinsic --
frequency dispersion of the current correlators at the scale set
by the voltage drop $V$ across the conductor. Since $V$
can vary in a wide range, this dependence is in general important
and should be taken into account while interpreting the experimental
results. In particular, in the absence of interactions the third
cumulant of the current operator ${\cal S}_3$ is fully determined by
the two parameters:
\begin{equation}
 \beta=\frac{\sum_n T_n(1-T_n)}{\sum_n T_n},\;\;
 \gamma=\frac{\sum_n T_n^2(1-T_n)}{\sum_n T_n} ,\label{bg}
\end{equation}
where $T_n$ represents the transmission of the $n$-th conducting
channel of our system. The cumulant ${\cal S}_3$ can be expressed
in the following general form
\begin{equation}
{\cal S}_3=(\beta -2\gamma F) e^2 \bar I.
\label{res}
\end{equation}
Here $\bar I$ is the average current through the conductor, $-e$ stands for
the electron charge, and $F$ is a universal function of frequencies,
voltage $V$ and temperature $T$ to be evaluated below. According to
Eq. (\ref{res}) the
frequency dispersion of ${\cal S}_3$ originates only from the term
proportional to the parameter $\gamma$ while the $\beta$-term is
dispersionless.

Though negligible for tunnel junctions $\gamma
\to 0$, the frequency dispersion of ${\cal S}_3$
may become important in other situations. For instance,
at $T\to 0$
one finds $F \to 1$ at frequencies much smaller than $eV$
while in the opposite high frequency limit one gets $F=0$ and, hence,
${\cal S}_3=\beta e^2 \bar I$ in the latter limit. To give some
numbers, for an important case of diffusive conductors one has $\beta
=1/3$ and $\beta -2\gamma =1/15$, i.e. in this case  the quantity
${\cal S}_3$ changes by the factor 5 depending on whether relevant
frequencies are below or above $eV$. For conductors with $\beta <
2\gamma$ even the sign of ${\cal S}_3$ will differ in these two limits.

In our analysis we will use the real time path integral formalism
developed for the systems of interacting fermions \cite{GZ97}. After
the standard Hubbard-Stratonovich decoupling of the interaction term
in the Hamiltonian one can exactly integrate out fermions and arrive
at the effective action $S$ which depends on the fluctuating fields
$V_{1,2}(t,\bm{r})$. Let us define
\begin{equation}
{\rm e}^{iS}={\rm Tr}\left[{\cal T}{\rm e}^{-i\int_0^t dt'{\bm H}_1(t')}
\hat{\bm \rho}_0 \tilde {\cal T}{\rm e}^{i\int_0^t dt'{\bm H}_2(t')}\right],
\label{eiS}
\end{equation}
with the trace taken over the fermionic variables. Here $\hat{\bm \rho}_0$ is
the initial $N-$particle density matrix of electrons,
\begin{eqnarray}
{\bm H}_{1,2}=\sum_{\sigma}\int d^3{\bm r}\,\hat\Psi^\dagger_\sigma({\bm r})
\hat H_{1,2}(t)\hat\Psi_\sigma({\bm r}),
\nonumber
\\
\hat H_{1,2}(t)=-\frac{\nabla^2}{2m}+U(\bm{r})-eV_{1,2}(t,\bm{r})
\label{H}
\end{eqnarray}
are the effective Hamiltonians on the forward and backward parts of the
Keldysh contour, $U(\bm{r})$ describes the static potential, ${\cal T}$ and
$\tilde{\cal T}$ are respectively the forward and backward time ordering
operators. Integrating out fermions in Eq. (\ref{eiS}), we obtain
$iS=iS_0+iS_{\rm em}$ where
\begin{equation}
iS_0=2\,{\rm Tr}\ln[1+(\hat u_2^{-1}\hat u_1-1)\hat\rho_0],
\label{S}
\end{equation}
$\hat u_{1,2}(t)={\cal T}\exp\left(-i\int_0^t dt' \hat H_{1,2}(t')\right)$
are the evolution operators pertaining to the Hamiltonians (\ref{H}),
$\hat\rho_0$ stands for the initial single-particle density matrix
and the term $S_{\rm em}$ accounts for the electromagnetic
contribution which may also include the effect of an external circuit.

In order to evaluate the evolution operators $\hat u_{1,2}$ it is necessary to
specify the model of a mesoscopic conductor. Here we will adopt the standard
model of a (comparatively short) coherent conductor placed in-between two bulk
metallic reservoirs. The electron dwell time $\tau_D$ is supposed to be
shorter than any relevant time scale in our problem. Energy and phase
relaxation times are, on the contrary, assumed to be long, i.e. inelastic
relaxation is allowed in the reservoirs but not inside the conductor. Under
these assumptions electron transport through the conductor can be described by
the energy independent scattering matrix
\begin{equation}
\hat S = \left( \begin{array}{cc} \hat r & \hat t'
\\ \hat t & \hat r'
\end{array}\right)
\label{sm}
\end{equation}
and the effective action (\ref{S}) can be expressed via the fluctuating phase
fields $\varphi_{1,2}$ which are in turn related to the jumps of the fields
$V_{1,2}$ across the scatterer as $\dot \varphi_{1,2}=e(V_{L1,2}-V_{R1,2})$,
where $V_{L,R}$ are fluctuating in time but constant in space fields in the
left and right reservoirs. We note that in this case the right hand side of
Eq. (\ref{eiS}) differs from the FCS generating functional introduced in Ref.
\onlinecite{KN2} only by a gauge transformation.

Within the above model the evolution
operators $\hat u_{1,2}$ were evaluated in Refs. \onlinecite{GZ00,GGZ}.
Combining these expressions with
Eq. (\ref{S}) after some algebra we find
\begin{widetext}
\begin{eqnarray}
iS_0=2{\rm Tr}\ln\left\{\hat 1\delta(x-y) +\theta(t-x)\theta(x)
\left[\begin{array}{cc} \hat t^\dagger\hat t({\rm e}^{i\varphi^-(x)}-1)
& 2i \hat t^\dagger\hat
r' \sin\frac{\varphi^-(x)}{2} \\  2i \hat r^{\prime \dagger}\hat t
\sin\frac{\varphi^-(x)}{2} & \hat t^{\prime \dagger}\hat t'( {\rm e}^{
-i\varphi^-(x)}-1)
\end{array}
\right]
\left[\begin{array}{c}
\rho_0(y-x){\rm e}^{i\frac{\varphi^+(x)-\varphi^+(y)}{2}}\hspace{0.9cm}  0
\\  0\hspace{0.9cm}  \rho_0(y-x){\rm e}^{i\frac{\varphi^+(y)-\varphi^+(x)}{2}}
\end{array}
\right] \right\}.
\label{S2}
\end{eqnarray}
\end{widetext}
Here we introduced $\varphi^+= (\varphi_{1} +\varphi_{2})/2$,
$\varphi^-= \varphi_{1} -\varphi_{2}$ and
\begin{equation}
\rho_0(x)=\int \frac{dE}{2\pi}\frac{e^{iEx}}{1+e^{E/T}}.
\end{equation}
Taking the trace in Eq. (\ref{S2}) implies convolution with respect to
internal time variables ($x$ and $y$). The total time span is
denoted by $t$.

Eq. (\ref{S2}) defines a formally exact effective action for a
coherent conductor described by an arbitrary energy independent
scattering matrix (\ref{sm}). This expression allows to fully
determine statistics of current fluctuations at arbitrary frequencies
and in the presence of interactions. In the case
of equilibrium fluctuations the formula (\ref{S2}) represents
a real time analogue of the effective action derived in Refs.
\onlinecite{SZ89,Naz} within the Matsubara technique. A formula
similar to Eq. (\ref{S2}) was also presented recently in
Ref. \onlinecite{KN}. In addition we note that,
provided the field $\varphi^-$ does not depend on time,
Eq. (\ref{S2}) coincides with the generating function considered,
e.g., in the problem of adiabatic
pumping through mesoscopic conductors \cite{AKL}.

Let us illustrate the relation between the effective action
(\ref{S2}) and the FCS generating function \cite{lev1}. For this
purpose we ($i$) disregard
interactions and set $\varphi^+(t)=eVt$ and ($ii$) suppress
fluctuations of $\varphi^-$ and set $\varphi^-=\rm{const}$. After these
simplifications from Eq. (\ref{S2}) we obtain
\begin{eqnarray}
iS_{FCS}=\frac{t}{\pi}\,{\rm Tr}\int
dE\, \ln\Big[1
 +\hat t^\dagger\hat t\, n_L(E)(1-n_R(E))
\nonumber\\ \times\,
({\rm e}^{i\varphi^-}-1)\label{fcs}
+ \hat t^\dagger\hat t\, n_R(E)(1-n_L(E))({\rm e}^{-i\varphi^-}-1)
\Big],
\label{FCS}
\end{eqnarray}
where $n_{L,R}(E)=1/[1+\exp[(E\pm eV/2)/T]]$.
Eq. (\ref{FCS}) is just the FCS generating function \cite{lev1}
which can be used to recover all cumulants of the current operator
in the zero frequency limit and in the absence of interactions.

The expression (\ref{S2}) is more general since it includes all possible
fluctuations of the phase fields $\varphi^{\pm}$.
For instance, if one allows for temporal
variations of $\varphi^-$, with the aid of Eq. (\ref{S2}) one can easily
describe the frequency dispersion of the current correlators. Below we
will illustrate this point by directly evaluating the third cumulant of
the current operator.

Within our model the current operator can be defined in a standard manner
\begin{equation}
\hat I(t)=\frac{e}{2}\frac{d}{dt}(\hat N_L(t)-\hat N_R(t)), \label{hatI}
\end{equation}
where $\hat N_{L(R)}$ is the total number of electrons in the left
(right) lead. Combining this definition with Eq. (\ref{eiS}) for
the expectation value one finds
$$
\langle \hat I(t)\rangle =-e\int {\cal D}\varphi^\pm \delta
S[\varphi^\pm]/\delta \varphi^-(t) e^{iS[\varphi^\pm]}.
$$
Similarly one can define higher moments of the current operator.
Here we consider the following correlation functions:
$\tilde{\cal S}_2= \langle\hat I(t_1)\hat I(t_2)+\hat I(t_2)\hat
I(t_1)\rangle /2$ and
\begin{eqnarray}
\tilde{\cal S}_3&=&\frac{1}{8}
\big\{\langle\hat I(t_1)\big({\cal T}\hat I(t_2)\hat I(t_3)\big)\rangle
+\langle\big(\tilde{\cal T}\hat I(t_2)\hat I(t_3)\big)\hat I(t_1)\rangle
\nonumber\\ &&
+\,\langle\hat I(t_2)\big({\cal T}\hat I(t_1)\hat I(t_3)\big)\rangle
+\langle\big(\tilde{\cal T}\hat I(t_1)\hat I(t_3)\big)\hat I(t_2)\rangle
\nonumber\\ &&
+\,\langle\hat I(t_3)\big({\cal T}\hat I(t_1)\hat I(t_2)\big)\rangle
+\langle\big(\tilde{\cal T}\hat I(t_1)\hat I(t_2)\big)\hat I(t_3)\rangle
\nonumber\\ &&
+\,\langle{\cal T}\hat I(t_1)\hat I(t_2)\hat I(t_3)\rangle
+\langle\tilde{\cal T}\hat I(t_1)\hat I(t_2)\hat I(t_3)\rangle\big\}.
\label{tri}
\end{eqnarray}
The correlation function $\tilde{\cal S}_2$ is important because
the symmetric combination of voltages $V^+$ can be viewed as a
classical, measurable, voltage \cite{KN2}. The
noise is deduced from the measurable product $V^+(t_1)V^+(t_2),$ which
is related to the symmetric correlator $\tilde{\cal S}_2$.
Similarly, the measured product
$V^+(t_1)V^+(t_2)V^+(t_3)$ is related to the correlation function of
the current operators $\tilde{\cal S}_3$ defined in
Eq. (\ref{tri}), see also Refs. \onlinecite{KN2,GG}.

Let us now disregard interaction effects. In this case
one should suppress fluctuations of $\varphi^{\pm}$
in the end of the calculation by setting
 $\varphi^+(\tau )=eV\tau$ and  $\varphi^-(\tau )\to 0$. Then for the noise
correlator one easily finds
\begin{eqnarray}
\tilde{\cal S}_2&=&(ie)^2\left.\frac{\delta^2 }{\delta\varphi^-(t_1)\delta
\varphi^-(t_2)}\,{\rm e}^{iS_0}\right|_{\varphi^-=0}
\label{dva}
\end{eqnarray}
and a similar expression is obtained for $\tilde{\cal S}_3$.
Of interest is the irreducible part of the correlator
$\tilde{\cal S}_3$ which reads
\begin{eqnarray}
{\cal S}_3&=&e^3\left.\frac{\delta^3 S_0}{\delta\varphi^-(t_1)
\delta\varphi^-(t_2)\delta\varphi^-(t_3)}\right|_{\varphi^-=0}
\label{S3} \\
&=&\tilde{\cal S}_3(t_1,t_2,t_3)-\langle\hat I(t_1)\rangle\tilde{\cal S}(t_2,t_3)
-\langle\hat I(t_2)\rangle\tilde{\cal S}(t_1,t_3)
\nonumber\\ &&
-\,\langle\hat I(t_3)\rangle\tilde{\cal S}(t_1,t_2)+2\langle\hat I(t_1)\rangle
\langle\hat I(t_2)\rangle\langle\hat I(t_3)\rangle.
\nonumber
\end{eqnarray}
It follows immediately that in order to evaluate the third current cumulant
in the absence of interactions it suffices to expand the exact effective
action (\ref{S2}) up to the third order in $\varphi^-$,
\begin{equation}
 iS_0[\varphi^\pm]=iS^{(1)}[\varphi^\pm]+iS^{(2)}[\varphi^\pm]+
iS^{(3)}[\varphi^\pm],
\end{equation}
keeping the full non-linearity in $\varphi^+$ in each of these terms.
The first two terms of this expansion were evaluated in
Ref. \onlinecite{GZ00}. Being combined with Eq. (\ref{dva}), the
term $S^{(2)}[\varphi^\pm]$
allows to recover the well known expression for the shot
noise spectrum \cite{bb}.
Proceeding further with the expansion in $\varphi^-$ for the term $iS^{(3)}$
one finds \cite{GGZ}
\begin{eqnarray}
&& iS^{(3)}[\varphi^\pm]=\frac{i\beta}{6 e^2 R}\int_0^{t}
d\tau(\varphi^-(\tau))^3\dot\varphi^+(\tau)\\&& -\frac{2\pi i\gamma}{3e^2
R}\int_0^{t} d\tau_1 \int_0^{t}d\tau_2
\int_0^{t}d\tau_3 \varphi^-(\tau_1)
\varphi^-(\tau_2)\varphi^-(\tau_3)\nonumber
\\&& \times f(\tau_2,\tau_1)f(\tau_3,\tau_2)f(\tau_1,\tau_3),\nonumber
\end{eqnarray}
where $1/R=(2e^2/h)\sum_nT_n$ is the scatterer conductance and
\begin{equation}
f(\tau_2,\tau_1)=\frac{T\sin\left[(\varphi^+(\tau_2) -\varphi^+(\tau_1))/2
\right]}{\sinh\left[ \pi T(\tau_2-\tau_1)\right]}.
\end{equation}
Let us substitute the above expressions into Eq. (\ref{S3}) and,
after taking derivatives over $\varphi^-$, set $\varphi^+(\tau )=eV\tau$ and
$\varphi^- \to 0$. This is sufficient provided the time differences
$|t_1-t_2|,$ $|t_1-t_3|$ exceed the charge relaxation time $\tau_{RC}$ and
provided $eV \ll 1/\tau_{RC}.$ The formula (\ref{S3}) immediately yields
\begin{eqnarray}
{\cal S}_3&=&\beta e^2\bar I\delta(t_1-t_2)\delta(t_1-t_3)\\ &&-\frac{4\pi e\gamma
}{R} f(t_2-t_1) f(t_3-t_2)f(t_1-t_3),\nonumber
\end{eqnarray}
where $f(\tau )=T\sin (eV\tau /2)/\sinh(\pi T\tau )$.
Performing the Fourier transformation
\begin{eqnarray}
{\cal S}_3(\omega_1,\omega_2)=\int d\tau_1 d\tau_2\,
{\rm e}^{i\omega_1\tau_1+ i\omega_2\tau_2}
{\cal S}_3(t_1,t_1-\tau_1,t_1-\tau_2)
\nonumber
\end{eqnarray}
we arrive at the final result
\begin{equation}
{\cal S}_3=\beta e^2 \bar I -2\gamma e^2 \bar I F(v,w_1,w_2),\label{eq1}
\end{equation}
where
\begin{equation}
F=\frac{\sinh^3(v/2)}{4v}\int_{-\infty}^\infty
 \frac{d\omega}{\chi (\omega )\chi (\omega-w_1) \chi (\omega+w_2)}.\label{funf}
\end{equation}
Here we defined $v=eV/T,\,w_{1,2}=\omega_{1,2}/2T$ and
\begin{equation}
\chi (\omega )=\cosh^2\omega+\sinh^2(v/4).
\label{chi}
\end{equation}
Eqs. (\ref{eq1})-(\ref{chi}) represent the main result of this paper.
They fully describe the third cumulant
of the current operator at voltages and frequencies smaller than
both $1/\tau_{RC}$ and $1/\tau_D$.

\begin{figure}
\includegraphics{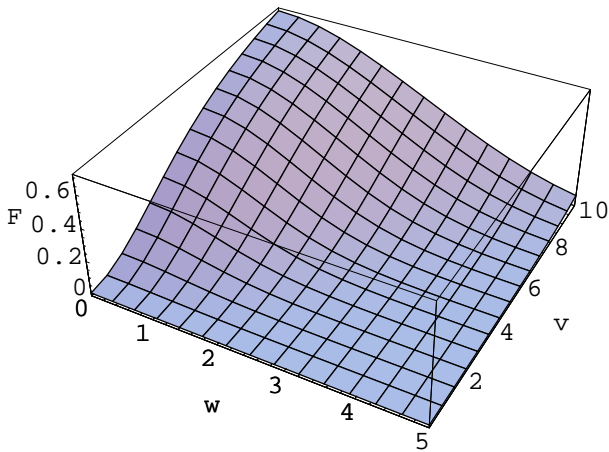}
\caption{The function $F(v,w,0)=F(v,w,-w)$, see Eq. (\ref{funf}).}
\end{figure}

Let us briefly analyze Eqs. (\ref{eq1})-(\ref{chi}) in various limits. For
$\omega_{1,2}=0$ we recover the well known result \cite{lev2}
\begin{equation}
F(v,0,0)=1+3\frac{1-(\sinh v/v)}{\cosh v-1},\label{levv}
\end{equation}
which in turn yields $F\to 1$ in the limit of large voltages $v \gg 1$.
Eq. (\ref{levv}) also holds for $w_{1,2}\ll v$.

In the limit $v\ll 1$ one finds
\begin{eqnarray}
&&F(v\ll 1,w,0)=F(v\ll 1,w,-w)=\label{as1}\\&&\frac{9\sinh w+\sinh
(3w)-12w\cosh w}{48\sinh^5 w}v^2\nonumber
\end{eqnarray}
and, similarly,
\begin{eqnarray}
&&F(v\ll 1,w,w)=\label{as2}\\&&\frac{\sinh
(4w)+4\sinh(2w)-8w\cosh(2w)-4w}{128\sinh^5 w\cosh^3 w }v^2\nonumber .
\end{eqnarray}
These equations demonstrate that at large frequencies $w \gg 1$  the function
$F$ decays exponentially with $w$. From Eqs. (\ref{as1}) and (\ref{as2})
we find respectively $F \propto v^2 e^{-2w}/3$ and $F \propto v^2 e^{-4w}$.

Finally let us turn to the most interesting limit of low temperatures, in which
case one always has $v,w_{1},w_2\gg 1$. Neglecting small corrections
$\sim 1/v,1/w$ we obtain
\begin{eqnarray}
F(v,w_1,w_2)= 1-2\left|\frac{w_{12}}{v}\right|,&&\text{if } 2|w_{12}|<|v|
\label{lt1}\\
F(v,w_1,w_2)= 0,&&\text{if}\;2|w_{12}|>|v|.
\label{lt2}
\end{eqnarray}
Here the value $w_{12}$ is defined differently depending on the sign of
the product
$w_1w_2$. For $w_1w_2>0$ we have $w_{12}=w_1+w_2$ while in the opposite case
$w_1w_2<0$ we define $w_{12}=\text{max}\,[|w_1|,|w_2|]$. We observe
that in both cases the function $F$ depends linearly on frequency and vanishes
as soon as $|w_{12}|$ exceeds $|v|/2$.

At arbitrary values of $v$, $w_1$ and $w_2$ the integral (\ref{funf}) can be
evaluated numerically. The corresponding result for the function $F(v,w,0)$ is
depicted in Fig. 1. The overall form of the function $F(v,w,w)$ is similar but
-- as compared to $F(v,w,0)$ -- it demonstrates a somewhat faster decay with
increasing frequency.

In conclusion, we have presented a general approach which allows to describe
statistics of current fluctuations in mesoscopic coherent conductors at
arbitrary frequencies and in the presence of interactions. Restricting
ourselves to the non-interacting case, we have analyzed frequency dispersion
of the third cumulant of the current operator. This dispersion was found
negligible only in the case of tunnel junctions, while in a general case it
turns out to be important in the frequency range comparable to $eV$. Similar
results are also expected for higher order cumulants of the current operator.

This work is part of the Kompetenznetz ``Funktionelle Nanostructuren''
supported by the Landestiftung Baden-W\"urttemberg gGmbH. One of us (A.V.G.)
acknowledges support from the Alexander von Humboldt Stiftung.

\end{document}